\documentclass[prl, aps, twocolumn, superscriptaddress, nofootinbib, tightenlines, nobibnotes, showpacs, groupaddress, 10pt]{revtex4-1}

\usepackage{natbib}
\usepackage{slashed}
\usepackage{graphicx}
\usepackage{subfigure}
\usepackage[usenames, dvipsnames]{color}
\usepackage{graphics}
\usepackage{hyperref}
\usepackage{bm}
\usepackage{comment}
\usepackage{amsmath}
\usepackage{color}\label{key}
\usepackage{amsfonts}
\hypersetup{colorlinks=true, linkcolor=blue, linktoc=page, citecolor=blue, urlcolor=blue}

\everymath{\displaystyle}

\begin{document}

\title{Conductance suppression due to two-stream instability in bilayer graphene}

\author{Vasco Pinh\~ao}
\affiliation{GoLP $\vert$ Instituto de Plasmas e Fus\~{a}o Nuclear, Instituto Superior T\'{e}cnico,
Universidade de Lisboa, 1049-001 Lisbon, Portugal}
\email{vasco.pinhao@tecnico.ulisboa.pt}

\author{Pedro Cosme}
\affiliation{GoLP $\vert$ Instituto de Plasmas e Fus\~{a}o Nuclear, Instituto Superior T\'{e}cnico,
Universidade de Lisboa, 1049-001 Lisbon, Portugal}

\author{Hugo Ter\c{c}as}
\affiliation{GoLP $\vert$ Instituto de Plasmas e Fus\~{a}o Nuclear, Instituto Superior T\'{e}cnico,
Universidade de Lisboa, 1049-001 Lisbon, Portugal}

\begin{abstract}
We investigate the electron-hole two-stream instability (or Coulomb drag) in intrinsic bilayer graphene in the hydrodynamic regime, accounting for the effects of temperature, initial drift velocity, magnetic field, and collisions. The threshold drift speed for the onset of instabilities is of the order of the thermal velocity of the carriers. We put in evidence an unprecedented purely electrostatic mechanism leading to current relaxation, giving rise to a well-defined dc longitudinal conductivity $\propto T^{3/2}$. Due to competition between electrostatic and collisional processes, two distinct transport regimes are identified. An analysis on the Hall conductivity revealed that the two-stream instability effects also correct the most recent results obtained within the linear response theory. 
\end{abstract}

\maketitle

\textit{Introduction.}--- The possibility of producing graphene samples of extreme purity has revealed an electronic regime that was realized theoretically in the 1960s \cite{RNGurzhi_1968}, but that always lacked experimental confirmation: the hydrodynamic regime. In this regime, electron-electron collisions dominate over scattering with disorder -- phonons and impurities --, and electronic dynamics can be described with classical fluid equations. This model has been extensively applied to the electron flow in monolayer graphene, whereas the hydrodynamics of graphene-based bilayer systems is not as well studied. In AB-stacked bilayer graphene, the electron and hole-like excitations have parabolic bands, thus a physical effective mass can be defined $m = 0.033$ $m_e$ ($m_e$ is the bare electron mass), allowing for a more natural hydrodynamic description, in contrast with the single-layer case, where a fictitious mass needs to be introduced in the models. Most of the work in this field concerns collective behavior away from charge neutrality (CN), partly because the hydrodynamic regime is usually only attainable at a high Fermi level and this is also the most common regime in technological applications. Such phenomena in electron-hole semiconductor plasmas have been considered in Refs. \cite{PhysRev.124.1387, PhysRevB.43.14009}, providing a linear response analysis of the two-stream instability in certain three and two-dimensional configurations. For typical parameters pertaining to the 2D structure discussed in \cite{PhysRevB.43.14009}, the instability was predicted under conditions that may not be experimentally realizable. In contrast, the high electron mobility in BLG offers a more favorable environment for observing these phenomena.\par
In this work, we investigate the electron-hole two-stream instability (or Coulomb drag) taking place at charge neutrality in bilayer graphene. We show that, below a critical temperature, there is a strong suppression of the conductivity due to the electrostatic turbulence generated in the instability process. Our findings may have important consequences in applications with bilayer graphene devices operating at the neutrality point.\par

\textit{Two-fluid model.}--- In light of the quadratic dispersion of the low-lying quasiparticles in BLG, $\epsilon(\mathbf{p}) = |\mathbf{p}|^2/2m$, the fluid equations take the familiar form for regular collisionless plasmas, with the exception that now the motion is confined to a plane. Thus, the system is described by the following set of conservation equations 
\begin{gather}
    \frac{\partial n}{\partial t} + \bm \nabla \cdot \mathbf{j} = 0,
    \label{cont_eq}\\
    \frac{\partial \mathbf{j}}{\partial t} \!+\! \bm\nabla \!\cdot\!\left(\frac{\mathbf{j}\otimes \mathbf{j}}{n} + \frac{P}{m}\mathbf{I}\right) = \frac{q}{m}(n\mathbf{E} + \mathbf{j}\times\mathbf{B}_0)+\frac{\mathbf{F}^\text{col}}{m},
    \label{vel_eq}\\
    \frac{\partial \mathbf{\epsilon}}{\partial t} + \frac{\partial}{\partial \mathbf{x}}\cdot\left((\epsilon +  P)\frac{\mathbf{j}}{n}\right) = q\mathbf{j}\cdot\mathbf{E}.
    \label{energ_eq}
\end{gather}
Here, $\mathbf{j}=n\mathbf{V}$ and $\epsilon$ are the current and energy densities, respectively. The constant $q = \pm e$ is the hole/electron charge and $P$ is the pressure of the electron fluid. The electric field $\mathbf{E} = \mathbf{E}_{\text{ind}} + \mathbf{E}_0$ is subdivided into a self-consistent portion $\mathbf{E}_{\text{ind}}$ and an eventual externally applied DC field $\mathbf{E}_0$, while the induced magnetic field is ignored because the flow is non-relativistic, i.e. $\mathbf{|V|}/c \ll 1$. Lastly, the $\mathbf{F}^\text{col}$ term encodes the processes of momentum dissipation or transfer between species.\par 
Under weak perturbations, the relaxation time approximation can be applied, leading to effective force collision force densities \eqref{relaxation_approx_e}. The intraspecies collision terms are absent because they conserve the total fluid momentum,
\begin{equation}
    \mathbf{F}^{\text{col}}_A = \mathbf{F}_{A, B} + \mathbf{F}_{A, \text{dis}} \approx -\frac{\mathbf{V}_A - \mathbf{V}_B}{\tau_{AB}} - \frac{\mathbf{V}_A}{\tau_{\text{dis},A}}.
    \label{relaxation_approx_e}
\end{equation}
In Ref. \cite{PhysRevB.101.035117}, the authors propose a phenomenological expression for the electron-hole scattering time that agrees satisfactorily with their numerical results in a temperature interval  $T \in \left[10 \hspace{3pt} \text{K}, 100 \hspace{3pt} \text{K}\right]$ close to the charge neutrality point, and so we adopt the same ansatz, $\tau_{e, h} = \tau_0(n_e + n_h)/n_h$, and  $\tau_{h, e} = \tau_0(n_e + n_h)/n_e$, where $\tau^{-1}_0 = 0.15 k_B T/\hbar$. Motivated by the experimental work \cite{Nam_2017}, we neglect the contribution from impurity collisions to the disorder scattering time, retaining only the electron-phonon interaction, whose experimentally motivated value is $\tau^{-1}_{\text{ph}} = 0.05 k_B T/\hbar$.\par
To close Eqs.\eqref{cont_eq}--\eqref{energ_eq} with a equation of state we resort to the local quasi-equilibrium hypothesis \cite{Lucas_2018}, we assume the two species to be distributed according to a drifting Fermi--Dirac distribution \cite{Crabb_2021, Lucas_2018}. The dynamic variables, in terms of temperature and chemical potential, can then be derived as   
\begin{gather}
    n_{e, h}(\mathbf{x}, t) = \frac{N_f m}{2\pi\beta_{e, h}\hbar^2}\ln{(1 + e^{\beta_{e, h}\mu_{e, h}})}\text{ and}
    \label{n_eq}
    \\
    \epsilon_{e, h}(\mathbf{x}, t)  = \frac{n_{e, h}m\mathbf{|V|}_{e, h}^2}{2}  -\frac{N_f m }{2\pi\beta_{e, h}^2\hbar^2}\text{Li}_2{(-e^{\beta_{e, h}\mu_{e, h}})}
    \label{e_eq}
\end{gather}
where $N_f = 4$ accounts for spin and valley degeneracy \cite{McCann_2013}, and $\text{Li}_2$ is the polylogarithm function of second order. The second term on the RHS of \eqref{e_eq} is the Fermi pressure of the fluid. To write an equation of state in terms of density, we invert the expression in \eqref{n_eq} to write the chemical potential as a function of density and plug the result into the pressure to obtain
\begin{equation}
    P_{e, h} = -\frac{N_f m}{2\pi\beta_{e, h}^2\hbar^2}\text{Li}_2\left[1 - \exp{\left(n_{e, h}\frac{2\pi\beta_{e, h}\hbar^2}{N_f m}\right)}\right].
    \label{eq_pressure}
\end{equation}
\indent\textit{Euler--Poisson system.}--- In what follows we ignore self-consistent magnetic fields and just consider the propagation of electrostatic waves in the system. Thus, with this simplification, it is enough to couple the two-fluid equations with Poisson's equation. In the long-wavelength limit where the mean-field description is valid, we do not expect the waves to resolve the interlayer distance $d \sim$ 1 \AA, thus we assume $kd \ll 1$ so that the Fourier transformed electrostatic potential becomes
\begin{equation}
    \phi(\mathbf{k}) = \frac{e}{2\varepsilon k}(n_h(\mathbf{k}) - n_e(\mathbf{k}))
    \label{phi_layers_eff_lim}
\end{equation}
In virtue of the range of temperatures reported in \cite{PhysRevB.101.035117}, from which the experimentally motivated scattering time estimates were extracted, we chose the typical temperature to be $T_0 = 50$ K. Then, the typical velocity, density, and energy scales can be defined in a consistent manner: $v_{50} \equiv \sqrt{k_B T_0/m}$; $n_{50} \equiv N_f m k_B T_0/(2\pi\hbar^2) \ln{2}$; $W_{50} \equiv k_B T_0$, respectively. Additionally, we introduce a typical length scale $L$, implicitly defining the time scale $L/v_{50}$. The scale invariance of the Euler equations implies that the two-fluid portion is independent of the length scale, but the source terms and coupling with the Poisson equation breaks this symmetry, as seen by expressing (\ref{phi_layers_eff_lim}) in such units
\begin{equation}
    \tilde{V} = \frac{e^2n_{50}L}{2\varepsilon W_{50}}\frac{(\tilde{n_h} - \tilde{n_e})}{\tilde{k}} = \frac{L}{\varepsilon_r\lambda_{\text{scr}}}\frac{(\tilde{n_h} - \tilde{n_e})}{\tilde{k}}
    \label{dimensionless_poisson}
\end{equation}
where $V \equiv e\phi$ is the potential energy. The tilded variables represent dimensionless quantities and $\lambda_{\text{scr}} = \frac{\pi}{\ln(2)}\frac{\varepsilon_0 \hbar^2}{e^2m} \approx 5.8 \text{ \AA}$ is the screening length, which sets the typical scale of charge imbalances in the plasma. The ratio $\lambda \equiv \lambda_{\text{scr}}/L$ controls the distance from quasi-neutrality, in the sense that as the domain size increases relative to $\lambda_{\text{scr}}$, the imbalances appear smaller and smaller at a macroscopic scale.\par
The standard time discretization of the two-fluid Euler--Poisson system consists of taking all the force terms implicitly \cite{Degond_2011, CRISPEL2007208} while leaving the fluxes explicit. We employ a Finite-Volume method \cite{ferziger2012computational, leveque2002finite, hirsch2007numerical} that preserves the global conservation property inherent to the Euler equations. In our scheme, the numerical fluxes were estimated a third-order MUSCL scheme \cite{toro2009riemann} coupled with a Min-mod slope limiter \cite{article_minmod} in a dimension-by-dimension manner and then computing the approximate Godunov flux \cite{toro2009riemann} at each boundary using the HLLC flux \cite{FLEISCHMANN2020109762}. Lastly, the time evolution is carried out with a third-order TVD Runge-Kutta method \cite{article_RK}.\par
\indent\textit{Two-stream instability.}--- The two-stream instability is a trademark instability in classical plasma physics, which is excited by placing two particle populations with a relative drift velocity in contact, e.g. a cold electron beam impinging on a steady plasma \cite{PhysRevLett.30.732}, or two counter-propagating beams. In an electron-hole plasma, both species have equal mass so, applying a stationary electric field will drive the electrons and holes in opposite directions at the same drift speed, thus also triggering this instability.
\begin{figure}[t!]
    \centering
    \includegraphics[height = 0.41\linewidth]{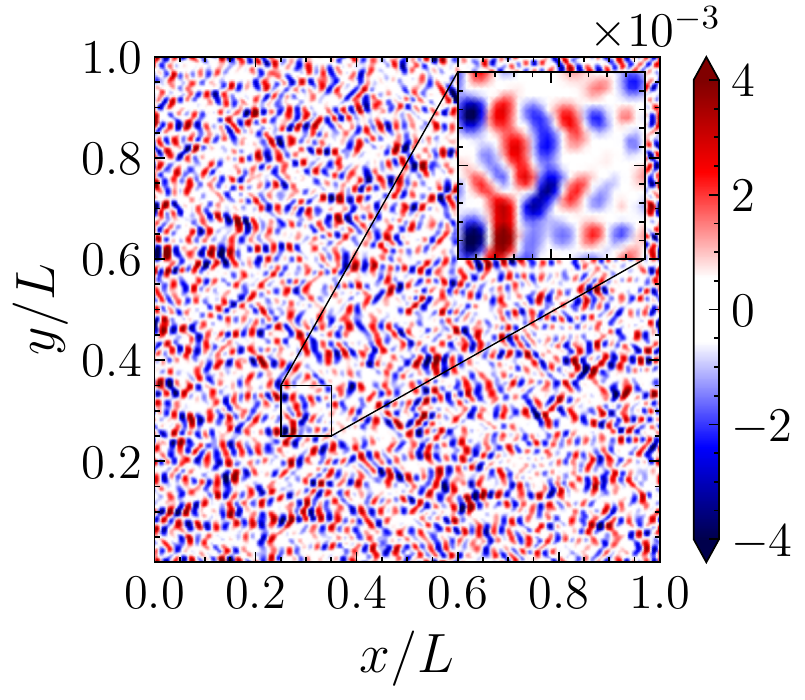}
    \includegraphics[height = 0.41\linewidth]{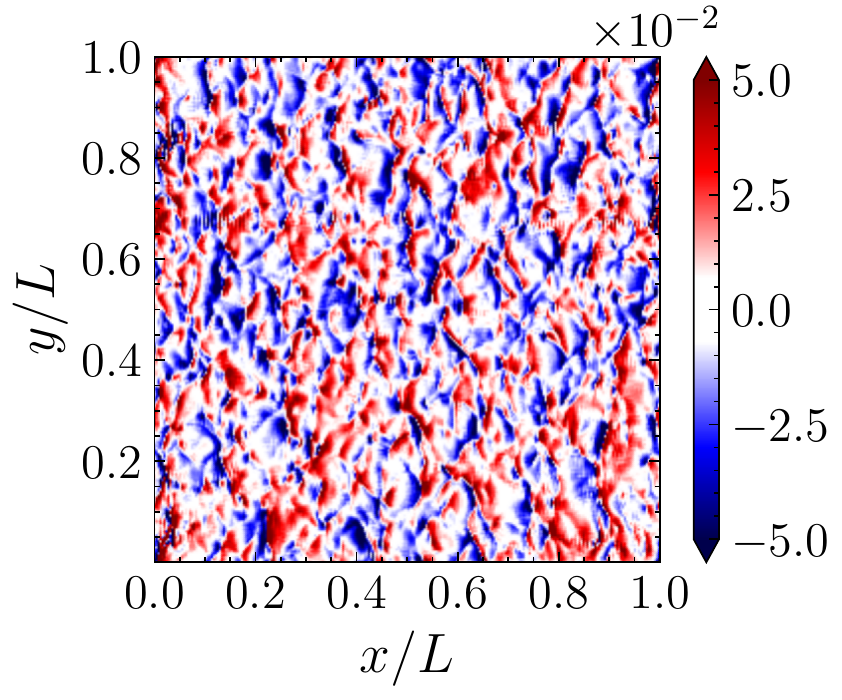}
    \caption{a) Charge imbalance density at the linear phase of the instability ($t = 0.016 L/v_{50}$), highlighting the formation of charge bunches in the inset; b) Charge imbalance density at the non-linear phase of the instability ($t = 0.18 L/v_{50}$). The simulations were performed on a 256$\times$256 grid with periodic boundaries for $v_0/v_{50}$ = 3.0 and $L$ = 1000 $\lambda_{\text{scr}}$.}
    \label{fig:1}
\end{figure}
A signature of this process is the formation of charge bunches, i.e. negative or positive charge accumulations of small width as shown in Fig.(\ref{fig:1}), whose amplitude grows exponentially in the linear phase. This phase ends when quasi-neutrality is restored as the bunches coalesce, after which the wave breaks. The ensuing saturation regime is characterized by a conversion of the wave energy into turbulent kinetic energy, hence requiring a non-linear analysis.\par 
Performing a linear analysis of the collisionless Euler--Poisson system for a configuration of counter-streaming electron and hole fluids with speed $v_0$ gives rise to a mode whose frequency can take on imaginary values, and therefore become unstable, under the condition
\begin{widetext}
\begin{equation}
    \omega^2  =
    -\sqrt{q_{\text{scr}}^2 v^4_{\text{th}}k^2 + 4 q_{\text{scr}} k_x^2 v^2_0  v^2_{\text{th}} k + 4 k_x^2 v^2_0 \left(v^2_{\text{th}}\ln (4) k^2 + \omega^2_c\right)} + q_{\text{scr}} v^2_{\text{th}} k + v^2_{\text{th}}\ln (4)k^2 + k_x^2v^2_0 +\omega^2_c < 0\label{inst_cond}
\end{equation}%
\end{widetext}
where $v_{\text{th}} = \sqrt{k_B T/m}$ is the thermal velocity, $q_{\text{scr}} = \lambda_{\text{scr}}^{-1}$ is the effective screening wavenumber, and $\omega_c = eB/m$ is the cyclotronic frequency.  We will admit that $\varepsilon = \varepsilon_0$ in all the graphical representations. If the condition \eqref{inst_cond} is fulfilled, the unstable mode has a vanishing real part, and so it is a stationary wave whose amplitude grows exponentially with a growth rate $\gamma \equiv \text{Im}(\omega)$.\par 
By solving $\gamma = 0$ with respect to $v_0$ and selecting the value of less magnitude, we find the minimum speed for which a mode is unstable to be given by
\begin{equation}
   v_{\text{min}}(\theta, T, \omega_c) \equiv \sqrt{\frac{\omega_c^2}{k_x^2} + \frac{v^2_{\text{th}}\ln(4)}{\cos^2{(\theta)}}}
    \label{cutoff_vel}
\end{equation}
where $\theta = \arctan{(k_y/k_x)}$ is the angle of $\mathbf{k}$ with respect to the $x$ axis. This minimum velocity is always non-zero for finite temperature and tends to the critical velocity $v_C \equiv \sqrt{v^2_{\text{th}}\ln(4)} \approx 1.117v_{\text{th}}$ as $\theta \rightarrow 0$, which equals the sound velocity associated with the linearized Fermi pressure (\ref{eq_pressure}). A qualitative microscopic argument \cite{PhysRevB.43.14009} can be put forward to explain this: for smaller drift speeds, the thermal spread is more pronounced, and so the particles in the fluid tend to add up incoherently in the density perturbations, diminishing the possibility of exciting an instability.\par
Looking at the cutoff wavenumber \eqref{k_cutoff} for $\omega_c$ = 0, we notice that for a fixed angle and temperature, its expression has a pole for $v_0 = v_C/\cos(\theta)$, where the width of the unstable band diverges.
\begin{equation}
    k_{\text{cutoff}} \equiv \frac{q_{\text{scr}} v_{\text{th}}^2\left|-2 v_C^2 + v_0^2\cos(2\theta) + v_0^2\right|}{\left(v_0^2\cos^2(\theta) - v_C^2\right)^2}
    \label{k_cutoff}
\end{equation}
As the projection of the wave vector along the initial velocity becomes smaller, the minimal velocity is adjusted so that, from the point of view of the particles of the wave, the drift speed remains in the proximity of the critical velocity, i.e. $\mathbf{k}\cdot\mathbf{v_0}/k \sim v_C$. Larger values of initial drift speed will have a well-defined cutoff wavenumber, which, unsurprisingly, shrinks as the projection $\mathbf{k}\cdot\mathbf{v_0}/k$ departs from the typical system parameters.\par
Since the temperature determines the number density of the species at CN and denser fluids favor the instability, there is a temperature range where the destabilizing effect of raising the density dominates over the stabilizing widening of the thermal spread. At temperatures such that $v_0 \sim v_{\text{th}}$, the latter contribution starts to play a more critical role and prevents instability.\par
The most unstable mode is the wave vector $\mathbf{k}$ = $(k_x, k_y)$ which minimizes the expression (\ref{inst_cond}) for a set of external parameters $(v_0, T, \omega_c)$. Its magnitude should not be larger than the thermal wavenumber and the inverse electron collision mean-free path, where quantum effects become important and a hydrodynamic description is inadequate. The screening length will in general be much smaller than this wavelength for the range of temperatures we considered.\par
%For illustration purposes we restricted $k$-space to a circle of radius $|\mathbf{k}| = 64\times 2\pi/L$ ($L = 1000\lambda_{\text{scr}}$) in Fig.(\ref{fig:2}). 
We find that the magnitude of the wave vector corresponding to the most unstable mode, at a constant temperature, is independent of $v_0$ and always equal to the maximum allowed value. 
\begin{figure}[t!]
    \centering
    \includegraphics[width = 0.215\textwidth]{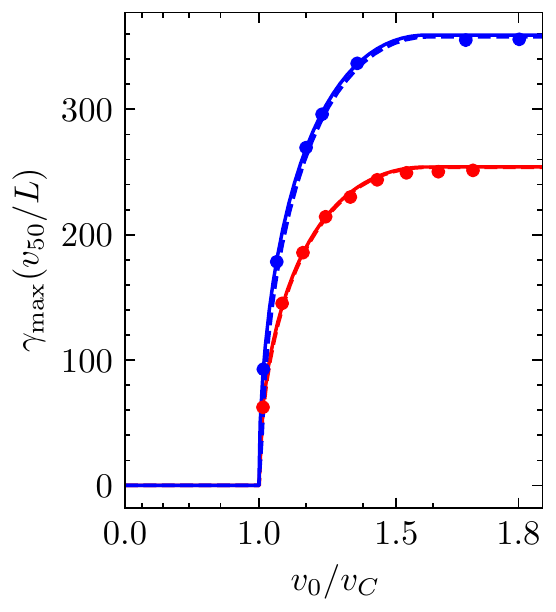}
    \includegraphics[width = 0.23\textwidth]{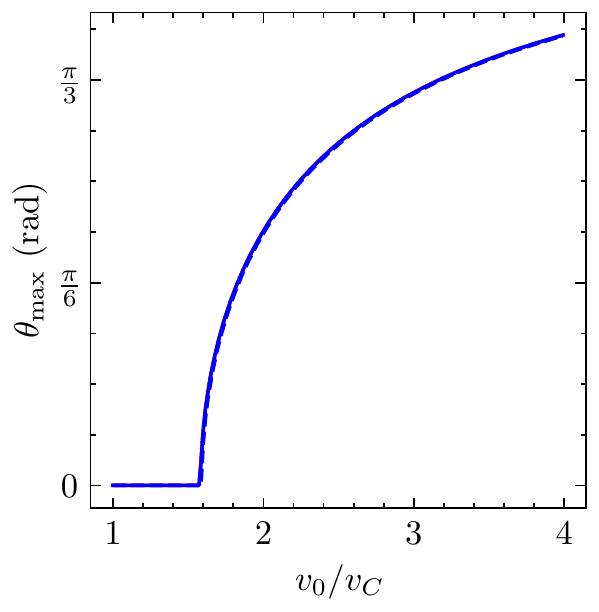}
    \caption{a) Maximum growth rate as a function of $v_0/v_C$ for $T = 50, 100$ K (Red, Blue) and $B$ = $0, 2, 5$ T (Solid, Blue Dashed, Red Dashed). The numerical estimates (Dots) for $B = 0$ are superimposed; b) Orientation of the wave vector corresponding to the most unstable mode. The same identification applies (note that the solid lines overlap).}
    \label{fig:2}
\end{figure}
The plot in Fig.(\ref{fig:2})-b) shows that, initially, this wave vector is aligned with the $x$ axis, but as $v_0$ increases, it starts rotating and tends asymptotically to $\theta = 90^\text{o}$. Interestingly, the mode directions for different temperatures collapse onto each other when plotted as a function of $v_0/v_C$, so the angle is a universal function of this quantity. Coincidentally, the rotation is also accompanied by a stabilization of the maximum growth rate, as illustrated in Fig.(\ref{fig:2})-a), meaning that $\gamma_{\text{max}}$ stops depending on $v_0$ when the most unstable wave vector develops a transverse component. Then, we set $k_y = 0$ in (\ref{inst_cond}) and minimize the expression with respect to $v_0$ to find the peak velocity
\begin{widetext}
\begin{equation}
        v_{\text{peak}} \equiv \frac{\sqrt{\left(k v^2_{\text{th}} (q_{\text{scr}} + k\ln(16)) + 2\omega_c^2\right) \left(k v^2_{\text{th}}(3q_{\text{scr}} + k\ln (16)) + 2 \omega_c^2\right)}}{2 k \sqrt{k v^2_{\text{th}} (q_{\text{scr}} +  k\ln(4)) + \omega_c^2}} \label{v_peak}
\end{equation}\end{widetext}
Hence, the maximum growth rate as a function of temperature, magnetic field, and wave vector magnitude is obtained by evaluating it at the peak velocity, in the long wavelength limit $k \ll q_{\text{scr}}$, it simplifies to
\begin{equation}
    \gamma_{\text{max}}(k \ll q_{\text{scr}}) = \frac{1}{2}v_{\text{th}}\sqrt{k q_{\text{scr}}}.
    \label{gamma_long_wave}
\end{equation}
For modes close to the instability threshold, the effect of collisions will eventually become noticeable but, for most temperatures, as the initial velocity is increased, there will still be available modes that are just slightly affected by collisions. This stops being the case when the maximum growth rate at a given temperature is of the order of the collision frequency, which occurs when
\begin{equation}
    T \geq T_{\text{col}} \equiv 16 \frac{\hbar^2 q_{\text{scr}}}{m k_B}k
    \label{t_col}
\end{equation}
We observe that only for very small wavenumbers do collisions start contributing significantly to the damping of the growing waves near room temperatures.\par 
Regarding the effect of magnetic field, for neutral BLG at finite temperatures, note that the validity of a semiclassical description, i.e. away from Landau levels \cite{PhysRevB.101.035117}, bounds the admissible magnetic field magnitude; for temperature $T = 50\sim100\,\text{K}$, it should not exceed $2\sim5\,\text{T}$. 
The plot in Fig.(\ref{fig:2})-a) confirms that the magnetic field damps the instability, although the effect is not very visible even for the maximally admissible field magnitudes. Alike collisions, only for very large wavelengths should we start observing significant damping. However, away from the weak-field limit, the linear theory stops being valid because we implicitly neglected the zeroth order term: $\mathbf{v}_0 \times \mathbf{B}_0$ in the linearized equations, but for stronger fields the cyclotronic motion induced in fluid significantly distorts the initial beam, translating into a loss of coherence which diminishes the average relative velocity between the two fluids and obviates the instability process. The angle $\theta_{\text{max}}$ is slightly translated to the right due to the shift in minimum velocity.\par
\begin{figure}[t!]
    \centering
    \includegraphics[width = 0.7\linewidth]{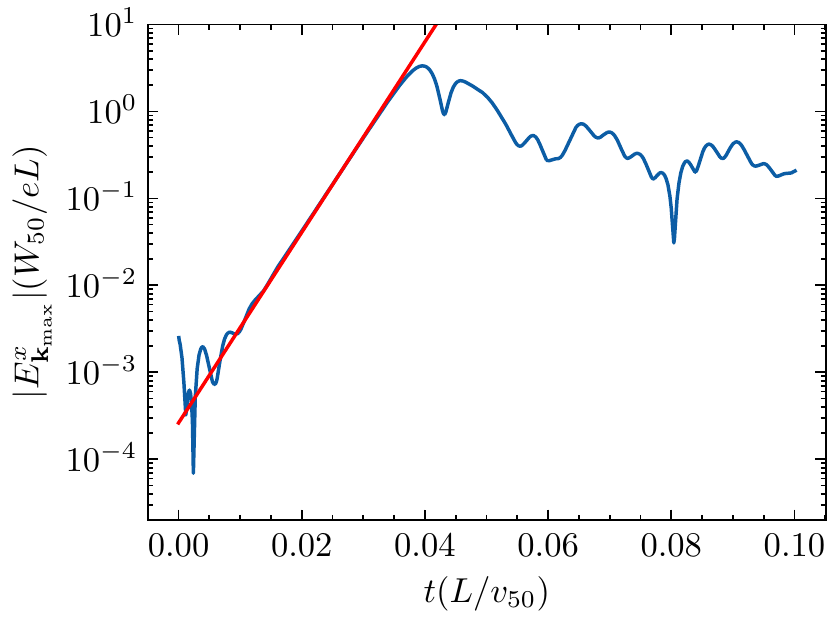}
    \caption{Time evolution of the most unstable Fourier coefficient of the electric field's $x$ component for $v_0/v_{50} = 2.0$, $T = 50$ K and $B = 0$ T. The simulation was performed on a 1024$\times$1024 grid with periodic boundaries.}
    \label{fig:3}
\end{figure}

From the longitudinal electric field nearest to the most unstable mode in Fourier space: $|E^x_{\mathbf{k}_{\text{max}}}|$, we estimated the corresponding growth rates Fig.(\ref{fig:3}) for the linear stage of the instability.
Such results plotted in Fig.(\ref{fig:2}) agree quite well with the theoretical predictions of the linear model for the selected values of initial velocity, thus validating the numerical method.\par
\indent\textit{Nonlinear Transport Properties.}--- Within the linear theory developed in \cite{PhysRevB.101.035117}, the DC longitudinal conductivity $\sigma_{xx} = \mathbf{J}_x(\omega = 0)/(\mathbf{E}_0)_x$ was found to be independent of temperature in the sole presence of Coulomb drag and phonon collisions, yielding
\begin{equation}
    \sigma^{\text{CN}}_{xx} = \frac{2 \ln(32)}{\pi} \frac{N_f e^2}{2\hbar} \approx 2.206 \frac{N_f e^2}{2\hbar}
    \label{sigma_CN}
\end{equation}
at CN. Throughout this section, the current we refer to is the total electric current density $\mathbf{J} = e(n_h\mathbf{V}_h - n_e\mathbf{V}_e)$. In the absence of a magnetic field and at CN, the linear theory predicts that a steady state is reached when the fluids acquire the speed
\begin{equation}
    v_h(t \rightarrow \infty) = -v_e(t \rightarrow \infty) = \frac{eE_0}{m}\frac{5\hbar}{k_B T}
    \label{limit_vel}
\end{equation}
Therefore, since there is a minimum fluid velocity (\ref{cutoff_vel}) to produce an instability, the asymptotic speed in (\ref{limit_vel}) must exceed that minimal speed to excite an unstable mode, which translates into the condition
\begin{equation}
     T \leq T_{\text{critical}} \equiv \left(\frac{25}{\ln(4)}\frac{e^2 \hbar^2}{m k_B^3}E_0^2\right)^{\frac{1}{3}} 
    \label{T_critical}
\end{equation}
Alternatively, under experimental conditions where the temperature is kept constant and not the electric field, it is possible to define a critical external field in an analogous manner
\begin{equation}
    E_0 \geq (E_0)_{\text{critical}} \equiv \frac{\sqrt{\ln(4)}}{5}\left(\frac{k_B^{3/2} m^{1/2}}{e\hbar}\right) T^{3/2}
    \label{E0_critical}
\end{equation}
which now constitutes a lower bound for instability.\par
The plasma modes amplified by the two-stream instability absorb energy from the translational degrees of freedom. In the turbulent regime, the size of the perturbations presents an additional dissipation mechanism in the form of non-linear wave-wave interactions, which compete with the growth due to the two-stream mechanism \cite{PhysRev.124.1387}. At saturation, the power supplied by the external electric field in the form of Joule heating $\mathbf{J}\cdot\mathbf{E}_0$ is appropriately redistributed among the plasma oscillations and the kinetic and thermal energy, producing an oscillating quasi-steady state.\par
\begin{figure}[t!]
    \centering
    \includegraphics[width = 0.23\textwidth]{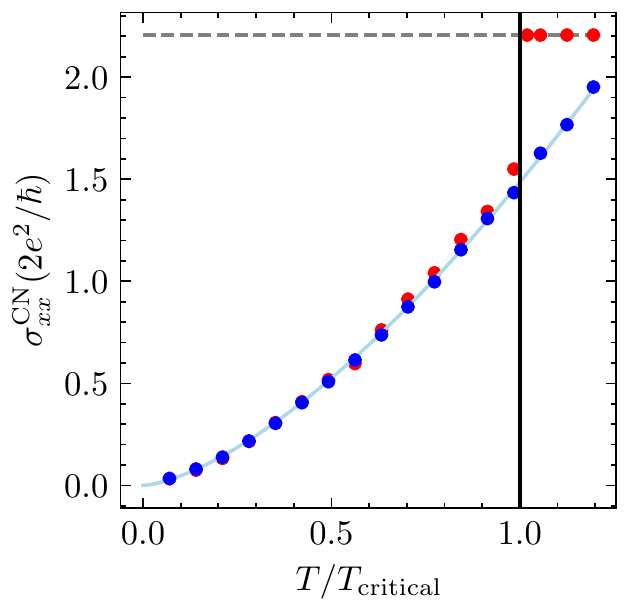}
    \includegraphics[width = 0.23\textwidth]{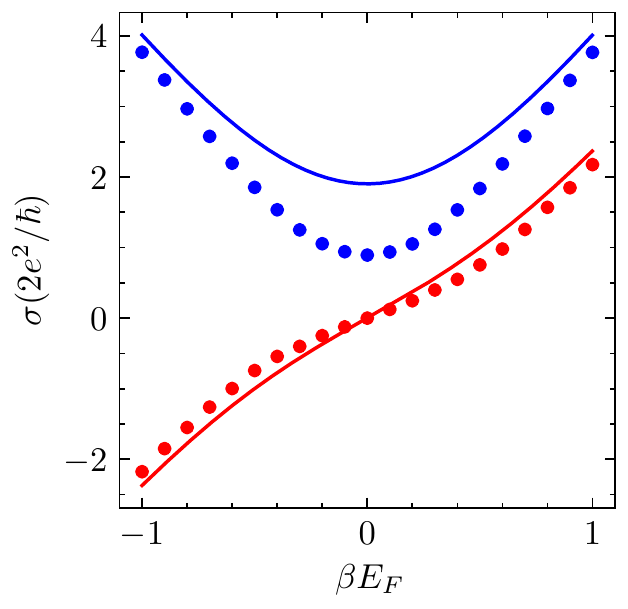}
    \caption{a) Longitudinal conductivity as a function of temperature in the collisionless (Blue Dots) and collisional (Red Dots) limit. The solid line is the best-fitted ansatz for the collisionless case and the dashed line represents the conductivity \eqref{sigma_CN}; b) Transverse (Red) and longitudinal (Blue) conductivities in the linear (Solid) and non-linear regimes (Dots) as a function of $\beta E_F$ for $T = 50$ K, $B \approx 0.05$ T. In both plots $E_0 \approx 74.4$ kV/m.}
    \label{fig:4}
\end{figure}

Motivated by the performed simulations, we propose a phenomenological ansatz (\ref{j_sat_ansatz}) for the saturation current
\begin{equation}
    \mathbf{J}^{\text{sat}}_x= A\mathbf{J}^{\text{min}}_x(T) = 2A en_0(T)\sqrt{\ln(4)\frac{k_B T}{m}} \propto T^{3/2}
    \label{j_sat_ansatz}
\end{equation}
in which $\mathbf{J}^{\text{min}}_x(T) \approx 2en_0(T)v_C$ is the approximate current magnitude before the instability, where $n_0$ is \eqref{n_eq} calculated at CN. This guess is justified both because the drift velocity amplitude saturates near $v_C$, hence the temperature dependence of the two currents should not vary appreciably, and the fact that, near the threshold for instability, the growth rate of the unstable modes is similar for each temperature, thus so should be the percentage of dissipated current. The fit shown in Fig.(\ref{fig:4})-a) provides $A(L = 100\lambda_{\text{scr}}) = 0.67$, meaning that about 33\% of the current at the onset of instability is dissipated. For better screening, there is greater energy transfer to the unstable waves, which establishes a lower saturation current: e.g., $A(L = 1000\lambda_{\text{scr}}) = 0.44$. In addition, stronger electric fields promote overshoots of the saturated velocity past $v_C$, enabling further current dissipation as more energetic modes are excited: e.g., the saturation current close to the critical electric field for $T$ = 50 K is $\sim 1.81$ in the units of the simulation, whereas for $E_0 = W_{50}/(e L) \approx 74.4$ kV/m it is $\sim 1.65$, exhibiting an obvious decline.\par
When collisions are introduced, two distinct regimes can be identified in Fig.(\ref{fig:4})-a). Before the critical temperature, $\sigma_{xx}$ follows the power law in \eqref{j_sat_ansatz}. After that value, we enter the linear regime, where the conductivity becomes constant with temperature, taking on the value predicted in (\ref{sigma_CN}).
Comparing the Hall conductivity Fig.(\ref{fig:4})-b) with the one obtained within the linear regime, we witness that there is a small suppression of the conductivity, but the overall shape remains the same. The same occurs for the longitudinal conductivity, but the suppression is more noticeable, particularly for small charge imbalances. Close to the threshold for instability, the transverse component of the unstable modes is null or much smaller than the longitudinal one, resulting in less dissipation of current due to the two-stream mechanism along the $y$ direction.\par
\indent\textit{Conclusions.}--- A hydrodynamic model for the flow of quasiparticles in BLG was developed and numerically implemented coupled with the Poisson equation. In the future, the numerical code should also incorporate viscous effects, neglected throughout our analysis, and that would stabilize the shocks driven by the convective terms.\par 
We proved the existence of a threshold drift speed for creating instabilities close to the thermal speed and a peak speed, after which the maximum growth rate saturates for a static cutoff wave number, which always coincides with the most unstable mode's wave vector norm. After the peak, this wave vector acquires a transverse component. Both an external magnetic field and collisions diminish the growth rate but, under typical parameters, their role only becomes apparent for very long wavelength modes. The theoretical predictions were confirmed numerically, validating the used scheme.\par
In the isothermal regime, two distinct regimes dominated, respectively, by electrostatic processes and collisions, were found.  The first exhibits a well-defined conductivity $\propto T^{3/2}$ at CN, contrasting with the temperature-independent value predicted by the linear theory. Away from CN, both $\sigma_{xx}$ and $\sigma_{xy}$ are suppressed relative to the linear case, although the first is more severely damped.\par
\textit{Acknowledgments.---} H.T. acknowledges Funda\c{c}\~{a}o da Ci\^{e}ncia e a Tecnologia (FCT-Portugal) through Contract No. CEECIND/00401/2018, and through the Project No. PTDC/FIS-OUT/3882/2020. P.C. acknowledges the funding provided by Fundação para a Ciência e a Tecnologia (FCT-Portugal) through the Grant No. PD/BD/150415/2019.

\bibliography{manuscript_v5.bib}% Produces the bibliography via BibTeX.

\end{document}